\documentclass[aps,superscriptaddress,prl,twocolumn,showpacs]{revtex4-1}

\usepackage{epsfig}
\usepackage{amsmath}
\usepackage{amssymb}
\usepackage{color}
\usepackage{bm}
\usepackage{dsfont}
\usepackage[normalem]{ulem} % for slash through text to be deleted
\usepackage{todonotes}
\usepackage{subfigure}

\usepackage{epstopdf}
\newcommand{\ket}[1]{\left|#1\right\rangle}
\newcommand{\ketc}[1]{\left|#1\right)}
\newcommand{\bra}[1]{\left\langle#1\right|}

\usepackage{verbatim}
\definecolor{purple}{RGB}{160,32,240}

\definecolor{seegroen}{rgb}{0,0.5,0.5}

\begin{document}
%=================================================================

%\title{Multidimensional Quantum Walks are Possible Using Classical Light and Linear Optical Setups}
\title{Implementation of multidimensional quantum walks using linear optics and classical light}
%-------------------------------------

\author{Sandeep K.~Goyal}
\email{sandeep.goyal@ucalgary.ca}
\affiliation{Institute of Quantum Science and Technology, University of Calgary, Calgary, Canada, T2N~1N4}
\affiliation{School of Chemistry and Physics, University of KwaZulu-Natal, Private Bag X54001, 4000 Durban, South Africa} 

\author{Filippus S.~Roux}
\affiliation{CSIR National Laser Centre, PO Box 395, Pretoria 0001, South Africa}
\affiliation{School of Physics, University of the Witwatersrand, Private Bag X3, Wits 2050, South Africa}

\author{Andrew Forbes}
\affiliation{School of Physics, University of the Witwatersrand, Private Bag X3, Wits 2050, South Africa}
\author{Thomas Konrad}
\affiliation{School of Chemistry and Physics, University of KwaZulu-Natal, Private Bag X54001, 4000 Durban, South Africa} 
\affiliation{National Institute for Theoretical Physics (NITheP), KwaZulu-Natal, Westville 4001, South Africa}

%===========================================
\begin{abstract}
Classical optics can be used to efficiently implement certain quantum information processing tasks with a high degree of control, for example, one-dimensional quantum walks through the space of orbital angular momentum of light directed by its polarization. To explore the potential of quantum information processing with classical light, we here suggest a method to realize $d$-dimensional quantum walks with classical optics---an important step towards robust implementation of certain quantum algorithms. In this scheme, different degrees of freedom of light, such as frequency, orbital angular momentum, and time bins, represent different directions for the walker while the coin to decide which direction the walker takes is realized by employing the polarization combined with different light paths. 
\end{abstract}

\maketitle

{\em Introduction.} A quantum algorithm is an ordered set of instructions given to a quantum computer to realize a particular processing of quantum information. However, due to the delicate nature of quantum systems, it is not easy to construct a large enough quantum computer  that is capable of implementing sufficiently complex quantum algorithms. Interestingly, a number of quantum algorithms, such as the Deutsch algorithm~\cite{Deutsch1985}, the Deutsch-Jozsa algorithm~\cite{Deutsch1992}, and Grover's search algorithm~\cite{Grover1996}, can be simulated and implemented with certain classical systems~\cite{Dyson2011} among them with classical states of light~\cite{Perez.et.al2015}.

%Quantum walks is a class of quantum processes {\bf that} can be used to perform quantum algorithms~\cite{Aharonov1993,Nayak2000,Ambainis2001,Kempe2003}. 
In principle, quantum walks can be used to realize quantum algorithms~\cite{Aharonov1993,Nayak2000,Ambainis2001,Kempe2003}. Algorithms such as quantum search~\cite{Berry2010,Shenvi2003} and graph isomorphism~\cite{Douglas2008,Gamble2010,Qiang2012} have already been demonstrated using quantum walks. Recently, it was proved that the most general quantum algorithms can be implemented using only quantum walks~\cite{Childs2009,Lovett2010}, thus making them universal for quantum computing. 
However, there is no efficient way known to implement an arbitrary algorithm using a  quantum walk with a single walker. 

Quantum walks can be realized with classical systems~\cite{Knight2003,Knight2003B,Roldan2005,Goyal2013,Schreiber2010,Schreiber2012}  that are simpler to prepare, control and measure than their quantum counterparts. Despite all these advantages, so far, only one-dimensional quantum walks have been implemented~\cite{Travaglione2002,Schmitz2009,Schreiber2010,Broome2010,Matjeschk2012,Peng_2014,Cardano2015}. There have been a few attempts to realize two-dimensional quantum walks experimentally~\cite{Schreiber2012}.  However, since the complexity of the realization schemes for two-dimensional quantum walks is much higher than that of their one-dimensional counterparts, it is unlikely that we see three- or higher-dimensional quantum walks merely based on quantum systems. For example, in Ref.~\cite{Schreiber2012}, Schreiber {\em et~al.} used a time-multiplexing technique with single photons to implement two-dimensional quantum walks. In this method, a two-dimensional lattice is mapped onto the one-dimensional discrete time line. 
 On the other, hand Roldan {\em et al.}~\cite{Roldan2005} presented a scheme to implement two-dimensional quantum walks where the propagation of the walker in the $x$ and $y$ directions is the displacement in the frequency domain corresponding to their respective linear polarizations.

 In this Rapid Communication, we present a scalable optical implementation scheme for a particular class of multidimensional quantum walk.  In this class a $d$-dimensional quantum walk is equivalent to $d$ one-dimensional quantum walks along with a $2d$-dimensional coin operator~\cite{Ambainis2005,Joye2012} (see also \footnote{{There are alternate ways of defining a $d$-dimensional quantum walk where the dimension of the coin space may vary. For example, the quantum walk defined in~\cite{Mackay2002} requires a $2^d$-dimensional coin space and  $N$-dimensional alternate quantum {walks~\cite{Roldan2013}} only require two-dimensional coin. Allowing for the possibility to move left, right or stay at the current position would even require a $3^d$-dimensional coin space. Quantum walks defined in these ways may not result in the similar dynamics and hence are not equivalent.}}). The scheme presented here requires only a linear optical setup and is capable of performing $d$-dimensional quantum walks on a regular lattice, where $d$ is the number of independent degrees of freedom (DoFs) of paraxial light. {It can be implemented using a single photon but also with coherent classical light. This is due to the analogy between the optics of single photons and coherent states prepared in the same light modes. In particular, the ingredients of quantum walks, namely, superposition, interference, and, indeed, a form of entanglement, are also present in classical optics~\cite{Forbes2014}. This analogy is used to simulate quantum walks for single photons with weak coherent states of light~\cite{Schreiber2010,Schreiber2012}. In Ref.~\cite{ Goyal2013}, we have argued that one-dimensional quantum walks with a single walker can be implemented using classical light. Here, we claim that classical light can also be used for multidimensional walks. Although possible, there are  disadvantages to using single photons to implement multidimensional quantum walks compared to classical light, as we discuss in the conclusions. 
 
 In general, a $d$-dimensional quantum walk results in multipartite entanglement between the different coordinates (as independent infinite-dimensional quantum systems) and the coin. 
 In the context of a realization with classical light this would amount to nonquantum entanglement between the different DoFs of the light beams. Here, we present a scheme where the classical counterpart of multipartite entanglement can be observed. 

To start with, consider a one-dimensional quantum walk with a two-dimensional coin. 
The dynamics of the walker can be characterized by two operators: (i) the conditional shift operator $S$ and (ii) the coin toss operator $U_c$. Let $\ket{\uparrow}$ and $\ket{\downarrow}$ be two orthogonal states of the coin. Then, the shift operator ${S}$ for the walker on the line can be defined as
\begin{align}
{S} &= {F}\otimes \ket{\uparrow}\bra{\uparrow} + F^\dagger\otimes \ket{\downarrow}\bra{\downarrow}.\label{shift-one}
\end{align}
Here, ${F} = \sum_j\ket{j+1}\bra{j}$ and ${F^\dagger}$ represent forward and backward propagators on the lattice, respectively. The coin toss operator $U_c$ shuffles the state of the coin after every shift, which can be defined by its action on the coin states of the walker as
\begin{align}
\ket{\uparrow} &\to \alpha \ket{\uparrow} + \beta\ket{\downarrow},\quad
\ket{\downarrow} \to -\beta^*\ket{\uparrow} + \alpha^*\ket{\downarrow},\label{hadamard}
\end{align} 
and it belongs to the group of $2\times 2$ unitary matrices SU$(2)$. Here, $\alpha$ and $\beta$ are two complex numbers such that $|\alpha|^2 +|\beta|^2 =1$. Thus, the quantum walk propagator reads
\begin{align}
Z = {S}(\mathds{1}\otimes U_c) =& {F}\otimes \ket{\uparrow}(\alpha\bra{\uparrow} -\beta^*\bra{\downarrow})\nonumber\\
& + F^\dagger\otimes \ket{\downarrow}(\beta\bra{\uparrow} +\alpha^*\bra{\downarrow}).
\end{align}
The combined operation $Z$ causes a toss of the coin state and a coin-state-dependent propagation of the walker on the lattice. 

One can extend the idea of one-dimensional quantum walks to multiple dimensions. The conditional shift operator [cp. Eq.\,\eqref{shift-one}] for quantum walks in $d$ dimensions reads 
\begin{align}
{S}^{(2d)} &= \sum_{n=1}^d \left({F}_n\otimes \ket{\uparrow_n}\bra{\uparrow_n} + {F}_n^\dagger \otimes \ket{\downarrow_n}\bra{\downarrow_n}\right),\label{shift-d}
\end{align}
where ${F}_n$ and ${F}_n^\dagger$ correspond to the forward and backward propagators for the $n$th direction and the sets $\{\ket{\uparrow_n},\,\ket{\downarrow_n}\}$ constitute the basis for a $2d$-dimensional coin space. The coin toss operator, on the other hand, is a unitary operator $U_c^{(2d)} \in U(2d)$ acting on the $2d$-dimensional state space of the coin. Thus, the quantum walk propagator $Z_{2d}$ for $d$-dimensional quantum walks reads
\begin{align}
Z_{2d} &=  {S}^{(2d)}(\mathds{1}\otimes U_c^{(2d)}).
\end{align}

Note that the conditional shift operator ${S}^{(2d)}$ defined in Eq.~\eqref{shift-d} can be rewritten as the sum of operators $S_n$ as
\begin{align}
S^{(2d)} &= \sum_{n=1}^d S_n,
\end{align}
where each of the operators ${S}_n$ represents a one-dimensional conditional shift operator identical to the one defined in Eq.~\eqref{shift-one}. Thus $d$-dimensional quantum walks can be realized by using $d$ independent, one-dimensional conditional shift operators assisted by a $2d$-dimensional coin toss operator. Therefore, the main challenges in realizing a $d$-dimensional quantum walk are (a) devising methods to implement $d$ one-dimensional independent quantum walks and (b) methods to realize $2d$-dimensional coin operations. Classical light beams possess a number of independent DoFs such as orbital angular momentum (OAM), time spacing, frequency spacing, and spatial displacements. There are several schemes to implement one- and two-dimensional quantum walks in these DoFs~\cite{Knight2003,Knight2003B,Roldan2005,Goyal2013,Schreiber2010,Broome2010}, thus establishing that classical light satisfy the first requirement (a). In the following, we demonstrate a scheme to realize an arbitrary $2d$-dimensional coin operation in classical light using only linear optics.

{\em Coin toss operator.} In our scheme, the $2d$-dimensional coin is realized by using $d$ copropagating light beams, each possessing a $2$-dimensional polarization space. In such a realization, the unitary operations on polarization states in individual beams and the beam splitters to mix different beams constitute the set of elementary operations. To construct an arbitrary $2d$-dimensional coin operation $U_c^{(2d)}$ based on linear optics, we need to decompose it into these elementary operations. 
The CS (``cosine-sine'') decomposition for unitary matrices~\cite{Paige1994,Dhand2015} enables us to achieve this. 

For example, using this decomposition we can write any unitary operator  $U_c^{(4)}$ acting on a  four-dimensional complex Hilbert space as
\begin{align}
U_c^{(4)} &= \left(\begin{array}{c|c}
L_1^\dagger & 0\\
\hline
0 & L_2^\dagger
\end{array}\right)
\mathds{U}^{(4)}
\left(\begin{array}{c|c}
R_1 & 0\\
\hline
0 & R_2\end{array}\right),\label{4d-decomposition}
\end{align}
with
\begin{align}
\mathds{U}^{(4)} = \left(\begin{array}{cc|cc}
\cos\theta_1 & 0 & -\sin\theta_1 & 0\\
0 & \cos\theta_2 & 0 & -\sin\theta_2\\
\hline
 \sin\theta_1 & 0 & \cos\theta_1 & 0\\ 
0 & \sin\theta_2 & 0 & \cos\theta_2
\end{array}\right).
\end{align}
\begin{figure}
\[ 
\begin{array}{c}
\includegraphics[width=6cm]{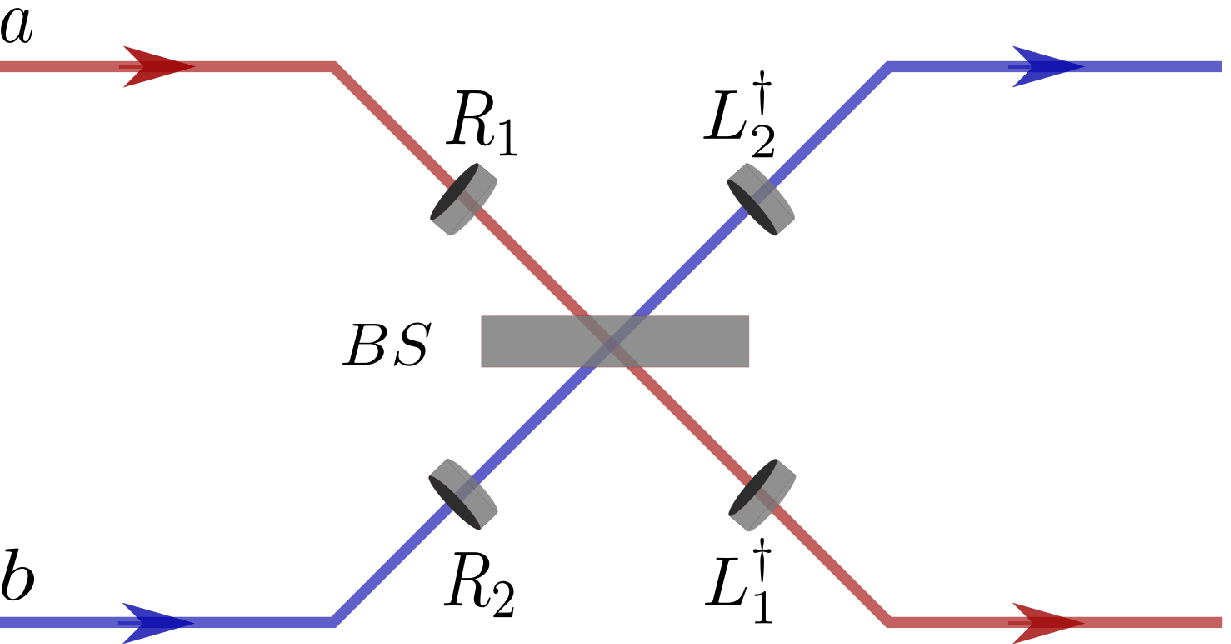}\\
\\
(a)\\
\\
 \includegraphics[width=6cm]{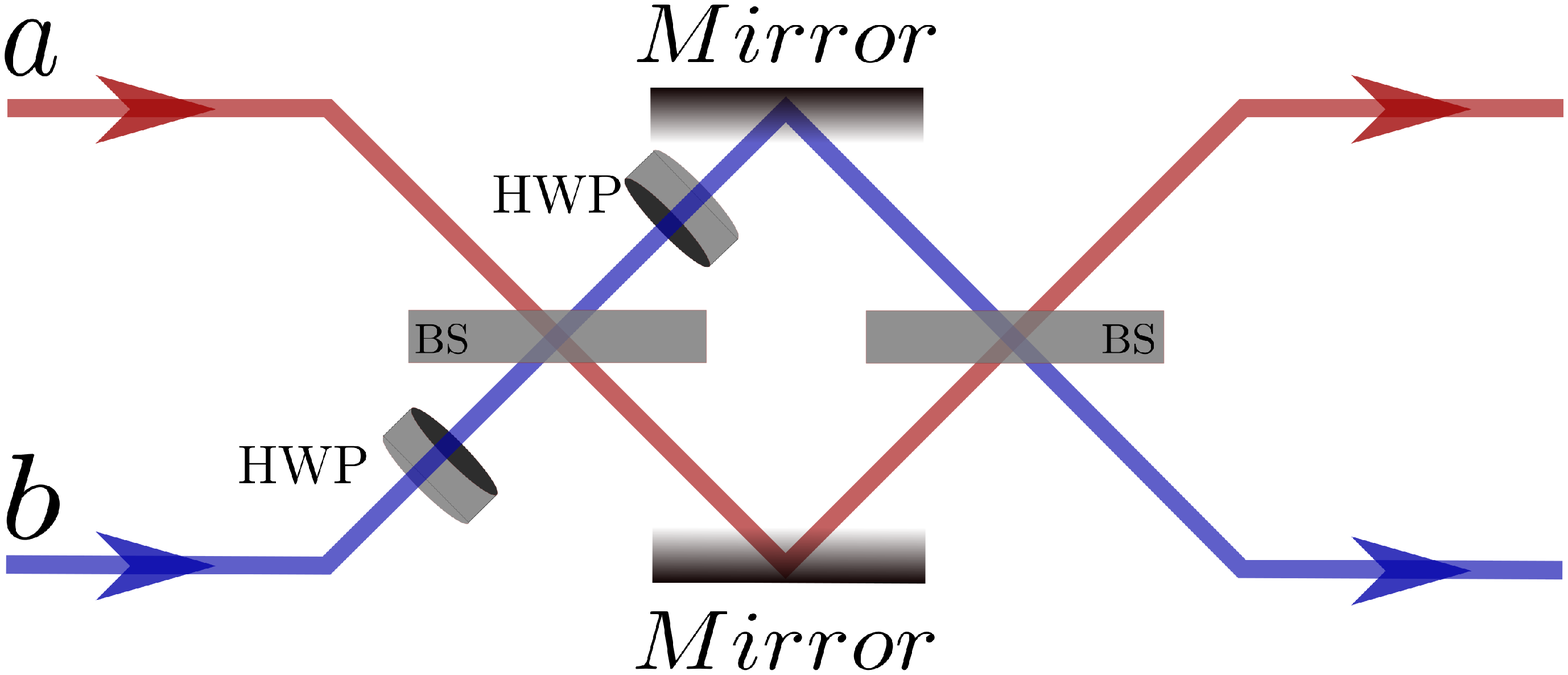} \\
\\
(b)
\end{array}\]
\caption{(Color online) (a) Diagram for the optical implementation of a four-dimensional unitary operation. Here, $BS$ stands for the elementary beam-splitter array implementing the operator $\mathds{U}^{(4)}$ in the CS decomposition (\ref{4d-decomposition}), which consists out of two beam splitters and two half-wave plates and shown in (b).}\label{coin-4d}
\end{figure}

\begin{figure}
\includegraphics[width=8.5cm]{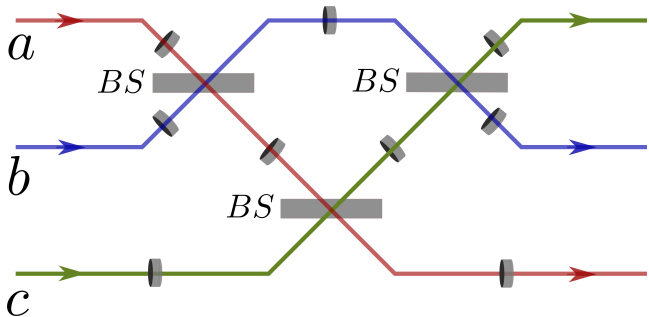}
\caption{(Color online) Any unitary coin toss operator for a quantum walk in three dimensions can be implemented by a combination of three elementary beam-splitter arrays $U^{(4)}$ combined with 9 unitary transforms of the polarization.}\label{coin-6d}
\end{figure}
The operators $R_1, R_2, L_1$, and $L_2$ are two-dimensional unitary matrices acting on the polarization in each beam. These operators can be realized by using a combination of quarter-wave-plates and half-wave-plates~\cite{Simon1990,Kok2010}, whereas the matrix $\mathds{U}^{(4)}$ is a four-dimensional unitary operator that is implemented by means of two beam splitters and two half-wave plates,
\begin{align}
\mathds{U}^{(4)}&=
\left(B(\theta_3)\otimes \mathds{1}_2\right)
\left(\mathds{1}_2\oplus \sigma_z\right)
\left(B(\theta_4)\otimes \mathds{1}_2\right)
\left(\mathds{1}_2\oplus \sigma_z\right),\label{eqn-09}
\end{align}
where
\begin{align}
B(\theta) = \left(\begin{array}{cc}
\cos\theta & -\sin\theta\\
 \sin\theta & \cos\theta 
\end{array}\right)
\end{align}
represents the beam-splitter matrix with transmittivity $\cos^2\theta$, $\theta_3 = (\theta_1 + \theta_2)/2$, and $\theta_4 = (\theta_1 - \theta_2)/2$, whereas $\sigma_z$ and $\mathds{1}_2$ represent the half-wave plates with the fast axis in the horizontal direction and a $2$-dimensional identity operator, respectively (see Fig.~\eqref{coin-4d}).

The decomposition of the four-dimensional coin operator $U_c^{(4)}$ given in Eq.~\eqref{4d-decomposition} can be extended to $2d$-dimensional coin operators~\cite{Dhand2015}. In that case the matrices $L_1$ and $R_1$ are two-dimensional unitary operators whereas, $L_2$ and $R_2$ are $(2d-2)$-dimensional unitary matrices. The matrix $\mathds{U}^{(2d)}$ can also be implemented by a polarization sensitive beam splitter acting on two beams and is of the form:
 \begin{align}
\mathds{U}^{(2d)} &= 
\left(\begin{array}{c|c}
\mathds{U}^{(4)} & 0\\
\hline
0 & \mathds{1}_{2d-4}
\end{array}\right).
\end{align}

The operators $L_2$ and $R_2$ can further be decomposed in $2$-dimensional and $(d-4)$-dimensional operators, and hence using CS decomposition repetitively we can create $U_C^{(2d)}$ by means of the elementary operations. In Fig.~\ref{coin-6d} we present a schematic representation of this decomposition for $d=3$. Thus, by using only the elementary operations we can perform an arbitrary $2d$-dimensional unitary operation. 

Once we have a $2d$-dimensional coin operator the rest of the scheme to implement the corresponding $d$-dimensional quantum walk is fairly straight forward. In Fig.\,\ref{fig-setup} we present the schematics for the realization of a three-dimensional quantum walk. In this figure we show a coin toss operator $U_c^{(6)}$ followed by three shift operators $S_1,\,S_2$, and $S_3$ in three independent paths named $a,\,b$ and $c$.
The shift operators $S_1,\,S_2$ and $S_3$ (in general) act on independent DoFs. For example, $S_1$ can be chosen as the shift operator acting on OAM states of the light beam~\cite{Goyal2013}. Likewise, the operators $S_2$ and $S_3$ could be implemented as shifts of the walker in time or in frequency domains~\cite{Schreiber2010,Schreiber2012,Roldan2005} and in position space. Hence, using our scheme, it is possible to implement a three-dimensional quantum walk with classical light.

{\em Quantum walk with Grover's coin.} As an example, in this section we present the scheme for quantum walks in two dimensions using the Grover coin~\cite{Mackay2002}, which, for two-dimensional quantum walks, is given by
\begin{align}
U^{(4)}_G &= \frac{1}{2}\left(\begin{array}{rrrr}
-1 & 1 & 1 & 1\\
1 & -1 & 1 & 1\\
1 & 1 & -1 & 1\\
1 & 1 & 1 & -1\end{array}\right).
\end{align}
Using the CS decomposition given in Eq.~\eqref{4d-decomposition}, this operator can be decomposed as
\begin{align}
U^{(4)}_G &= \left(\begin{array}{c|c}
-H & 0\\
\hline
0 & H \end{array}\right) \mathds{U}_G  
\left(\begin{array}{c|c}
H & 0\\
\hline
0 &\sigma_z H \end{array}\right),
\end{align}
where $H$ and $\mathds{U}_G$ read
\begin{align}
H&= \frac{1}{\sqrt{2}}\left(\begin{array}{rr}
1 & 1\\
1 & -1 \end{array}\right),\quad\mathds{U}_G = \left(\begin{array}{cccc}
0 & 0 & -1 & 0 \\
0 & 1 & 0 & 0 \\
1 & 0 & 0 & 0 \\
0 & 0 & 0 & 1\end{array}\right).
\end{align}
The operator $H$ can be realized by a half-wave plate with the fast axis at an angle $\pi/8$ with respect to the horizontal axis, whereas the operator $\mathds{U}_G$ can be realized using the scheme given in Eq.\,\eqref{eqn-09} with $\theta_3 = \theta_4 = \pi/4$.

If we choose OAM and the time bins as two independent DoFs, then any state vector of the walker can be written as a generalized Jones vector (path-polarization vector~\cite{Jones1941a, Hurwitz1941, Jones1941b, Jones1942}),
\begin{align}
\ket{\psi} &= \left(\begin{array}{r}
\sum_{\ell,t_n}\alpha_{\ell,t_n}E(t_n)\exp(i\ell\phi)\\
\sum_{\ell,t_n}\beta_{\ell,t_n}E(t_n)\exp(i\ell\phi)\\
\sum_{\ell,t_n}\gamma_{\ell,t_n}E(t_n)\exp(i\ell\phi)\\
\sum_{\ell,t_n}\delta_{\ell,t_n}E(t_n)\exp(i\ell\phi)\end{array}\right),
\end{align}
where $\alpha,\,\beta,\,\gamma,\,\delta$ are complex amplitudes of horizontally and vertically polarized light in the first and the second beam such that $\sum_{\ell,t_n}\left(|\alpha_{\ell,t_n}|^2+|\beta_{\ell,t_n}|^2+|\gamma_{\ell,t_n}|^2+|\delta_{\ell,t_n}|^2\right) =1$, whereas $E(t_n)$ denotes the temporal shape of the laser pulse centered around time $t=t_n$. It can be chosen to be a Gaussian pulse with temporal width $\tau$ and can be written as
  \begin{align}
E(t_n) &= \exp\left(-\frac{(t-t_n)^2}{2\tau^2}\right).
  \end{align}

In this notation the shift operators read
\begin{align}
S_1 &= \left(\begin{array}{rrrr}
e^{i\phi} &0 & 0 & 0\\
0 & e^{-i\phi} &0 &0\\
0& 0& 0& 0\\
0& 0& 0& 0\end{array}\right),
S_2 = \left(\begin{array}{rrrr}
0 &0 & 0 & 0\\
0 & 0 &0 &0\\
0& 0& f(\Delta t)& 0\\
0& 0& 0& f(-\Delta t)
\end{array}\right).
\end{align}
Here, $\Delta t$ is the step length in the time domain and $f(\Delta t)$ represents an operator  such that 
  \begin{align}
  f(\Delta t) E(t_n) = E(t_n + \Delta t),
  \end{align}
which can be realized by a Mach-Zehnder interferometer with unequal path lengths.

If the initial value of the OAM of the walker is zero, his initial time is $t_0$ and the light is initially travelling in the first beam with horizontal polarization, then the state of the walker reads
\begin{align}
\ket{\psi_0} &= E(t_0)\left(\begin{array}{r}
1\\
0\\0\\0\end{array}\right). 
\end{align}
After a single step of the quantum walk the state transforms to
\begin{align}
 \ket{\psi_1} = SU^{(4)}_G \ket{\psi_0} = \frac{1}{2}\left(\begin{array}{c} -E(t_0)e^{i\phi} \\E(t_0)e^{-i\phi}\\ E(t_0+ \Delta t)\\E(t_0- \Delta t)\end{array}\right).\label{GQW-1step}
\end{align}
Thus, by repeating this operation, one can perform the Grover quantum walks in two dimensions.

\begin{figure}
\includegraphics[width=8.5cm]{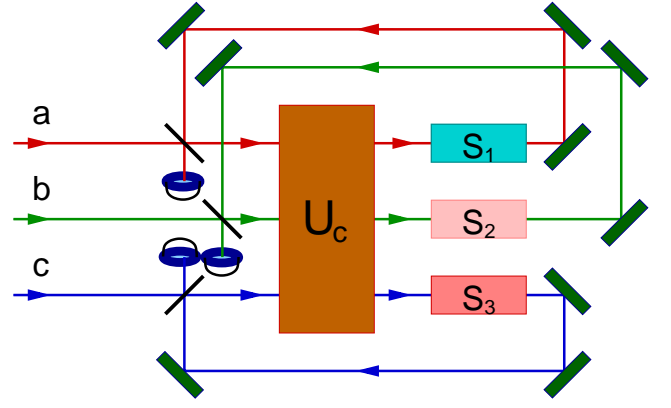}
\caption{(Color online) Schematic diagram for the setup for a $3$-dimensional quantum walk. Here, $a$, $b$ and $c$ denote the three paths for the light beams, $U_c$ is the $6$ dimensional coin flip operator, and $S_1$, $S_2$, and $S_3$ are the three shift operators acting on OAM, time, and the position space, respectively. }\label{fig-setup}
\end{figure}

{\em Multidegree nonquantum entanglement.} The scheme presented in this Rapid Communication  enables a multidimensional quantum walk in different DoFs of light. The classical nature of the scheme provides the usual benefits such as immunity against disturbance, it is technically simple to realize, and measurements can track the spread of the walk in real time. Beside these advantages, our scheme provides a platform to realize the classical counterpart of multipartite quantum entanglement, i.e., multidegree entanglement in classical light beams. This is due to the entangling property of quantum walk evolution, which tends to entangle different coordinates of the walker.  In our scheme the independent DoFs of the light beam serve as different coordinates of the walker. Thus, the quantum walk evolution results in entanglement between these DoFs of the classical light beam. However, since the whole setup is classical in nature, the entanglement in question is {\em not } quantum~\cite{Goyal2013, Simon2010}.

To understand this in more detail, let us consider again the Grover quantum walk. After a single step of the quantum walk the state of the walker [cf. Eq.\,\eqref{GQW-1step}] can be rewritten as
\begin{align}
\ketc{\psi_1}=& \frac{1}{2}\big(-\ketc{1}\ketc{t_0}\ketc{h_1} + \ketc{-1}\ketc{t_0}\ketc{v_1} \nonumber\\
&+\ketc{0}\ketc{t_0+\Delta t}\ketc{h_2} +\ketc{0}\ketc{t_0-\Delta t}\ketc{v_2}\big),\label{GQW_1step2}
\end{align}
where $h_1,v_1,h_2,v_2$ represent the horizontal and vertical polarization modes in the first and the second beam and $\ketc{1}\ketc{t_0}\ketc{h_1} = [E(t_0)e^{i\phi}~ 0~ 0~ 0]^T$, etc. It is clear from Eq.\,\eqref{GQW_1step2} that the state $\ketc{\psi_1}$ cannot be written as the product of OAM modes, time bins, and the polarization, hence it represents an entangled state. However, since the whole setup is classical, the entanglement is nonquantum.

In conclusion, we have presented a scheme to realize multidimensional quantum walks using the different DoFs of a light beam. The scheme uses only bright classical light  and is therefore very robust. Bright laser pulses are considerably simpler to produce, to process, and to detect than single photons. This makes it possible to perform a quantum walk on a large lattice. With the state-of-the-art technology it is possible to measure $100$ OAM states of light~\cite{Dudley2013,Goyal2013}. Similarly, one-dimensional quantum walks can be performed for $50$ steps in time bins~\cite{Schreiber2010,Schreiber2012} and for $10$ steps in spatial modes~\cite{Peng_2014}. Thus, using our scheme and these available methods to perform one-dimensional quantum walks, we can perform  between $10$ and $20$ steps three-dimensional quantum walks. Moreover, in principle, losses can be compensated by amplification unlike in the quantum regime of light, resulting in many steps possible in the  multidimensional quantum walk.  One of its most intriguing aspects is the occurrence of the classical counterpart of quantum multipartite entanglement, i.e., entanglement between more than two DoFs.  Since quantum walks are not possible without the entanglement between different directions along with coin states, multi-degree entanglement is essential for classical implementation of the multidimensional quantum walks. Devising the use of this particular type of entanglement will motivate our future studies.

\end{document}